# Micro-scale investigations of temperature-dependent Microbial-Induced Carbonate Precipitation (MICP) in the temperature range 4-50 °C

Yuze Wang[1,2], Yong Wang[1], Kenichi Soga[3], Jason T. DeJong[4], Alexandre J. Kabla[5]

**ABSTRACT** Microbially-Induced Carbonate Precipitation (MICP) involves a series of bio-geochemical reactions whereby microbes alter the surrounding aqueous environment and induce calcium carbonate precipitation. MICP has a broad range of applications, including *in-situ* soil stabilization. However, the reliability of this process is dependent on a number of environmental conditions. In particular, bacterial growth, bacterial activity, and precipitation kinetics all depend on temperature. Batch test and microfluidic chip experiments were performed in this study to investigate the effects of temperature on bacterial density and activity and the MICP processes occurring at different temperatures (4-50°C). Spatial and temporal variations in the formation and development of calcium carbonate precipitates, including their amount, type, growth rate, formation, and deformation characteristics, were monitored. Results show that different types of calcium carbonate precipitates with varying sizes and quantities were produced by varying the temperature. Low temperature (4°C) did not reduce bacterial activity, but limited the final amount of cementation; low temperature reduced bacterial growth and attachment ratio, as well as calcium carbonate precipitation rate. High temperature (50°C) conditions significantly reduced bacterial activity within a short period of time, whilst a repeated injection of bacteria before every two injections of cementation solution increased the final amount of cementation. The findings made from this paper provide insight into how MICP processes vary across a range of temperatures and could be valuable for optimising the MICP process for different applications.

**INTRODUCTION**

Microbially-Induced Carbonate Precipitation (MICP), particularly via the urea-hydrolysis pathway, has been extensively studied for its potential subsurface applications including permeability reduction of porous and fractured medium (Nemati and Voordouw 2003; Al Qabany and Soga 2013), soil stabilization (Whiffin et al. 2007; DeJong et al. 2010; van Paassen et al. 2010; Cheng and Cord-Ruwisch 2014) and environmental remediation (Achal et al. 2011; Jiang et al. 2019; Peng et al. 2020). During MICP treatment, bacteria with ureolytic activity produce a urease enzyme that catalyses the hydrolysis of urea (Eq. 1). The addition of calcium ($Ca^{2+}$) to this system then induces precipitation of calcium carbonate ($CaCO_3$) as $CO_3^{2-}$

[1] PhD, Department of Ocean Science and Engineering, Southern University of Science and Technology, Shenzhen, 518055, China；Southern Marine Science and Engineering Guangdong Laboratory (Guangzhou), Shenzhen, 518055, China (corresponding author). ORCID: 0000-0003-3085-5299. E-mail: wangyz@sustech.edu.cn

[2] Department of Ocean Science and Engineering, Southern University of Science and Technology, Shenzhen, 518055, China

[3] PhD, FREng, FICE, Department of Civil and Environmental Engineering, University of California, Berkeley, CA 94720, United States; 0000-0001-5418-7892

[4] Department of Civil and Environmental Engineering, University of California, Davis, CA 95616, United States; ORCID: 0000-0002-9809-955X

[5] PhD, Department of Engineering, University of Cambridge, Cambridge, CB2 1PZ, United Kingdom; ORCID: 0000-0002-0280-3531

Full contact details of corresponding author.
Yuze Wang, E-mail: wangyz@sustech.edu.cn; Department of Ocean Science and Engineering, Southern University of Science and Technology, Shenzhen, China ions react with $Ca^{2+}$ (Eq. 2)

$$CO(NH_2)_2 + 2H_2O \xrightarrow{Urease} 2NH_4^+ + CO_3^{2-} \quad (1)$$

$$Ca^{2+} + CO_3^{2-} \rightarrow CaCO_3(s) \quad (2)$$

The precipitated calcium carbonates produced by these *in-situ* biochemical reactions have multiple functions in porous soil media, such as reducing pore space, bonding soil particles, and co-precipitating with heavy metals or radionuclides.

The reactions not only depend on the quantity and activity of bacteria, and the content and concentration of cementation solution, but also on the environmental parameters such as temperature, pH, saturation conditions, and salinity (Stocks-Fischer et al. 1999; Ferris et al. 2004; Dupraz et al. 2009; Mortensen et al. 2011). Therefore, understanding the effects of environmental factors on the biochemical reactions is needed to predict and control the development of $CaCO_3$ crystals which affects the engineering properties of MICP-treated soils.

Temperature is one of the most important environmental factors to consider. In addition to affecting bacterial growth, death, and activity which consequently influences the rate of urea hydrolysis (Eq.1), temperature also affects the rate of precipitation (Eq.2) and the amount and morphology of the $CaCO_3$ crystals formed. Most previous MICP studies were conducted at room temperatures (20-25°C), whereas the temperature in the engineering field varies (Green and Harding 1980; Changnon 1999; Chow et al. 2011). For instance, if MICP is used to mitigate sand production during methane gas extraction from hydrate-bearing soils, the operational temperature could be between 0°C and 12°C (Peltzer and Brewer 2000; Koh et al. 2002; Sagidullin et al. 2019); if MICP is applied to seal the fractures of geoformation for $CO_2$ storage, the temperature in the deep subsurface could be higher than 31°C (Mitchell et al. 2009); when MICP is used to stabilize subsurface soils in arid regions in hot summer, the soil temperature may be as high as 50°C (Gao et al. 2007; Wu and Zhang 2014).

Some of the previously published studies have examined the effect of temperature on crystals characteristics, carbonate amount and the strength improvement of MICP-treated soils. Cheng et al. (2017) conducted soil column experiments and showed that compared with 25°C, either low (4°C) or high (50°C) temperature results in the formation of relatively smaller calcium carbonate crystals compared to the size of crystals produced at 25°C. The reasons for the production of smaller calcium carbonate crystals are assumed to be because of low bacterial activity at 4°C, and larger numbers of crystal nucleation produced at 50°C (Cheng et al., 2017). Sun et al. (2018) suggested that, in the range of 15 to 30°C, higher temperature can precipitate more calcium carbonate within 48 hours, because bacterial reproduction and enzyme activity increase at higher temperatures which results in increased precipitation efficiency. By contrast, Peng et al. (2019) conducted batch tests and reported that the urea hydrolysis activity of bacterial suspensions increased to a peak and then decreased over time within 400 hours, and the higher the temperature, the faster the activity decreased. In addition, by conducting soil column experiments, Peng et al. (2019) found that in the temperature range of 10-30°C, higher temperatures resulted in less precipitation of $CaCO_3$.

In the light of these results, it is necessary to further investigate the effect of temperature on bacterial activity and the dynamics of $CaCO_3$ precipitation. Earlier studies have shown that bacterial density, which may change due to their *in-situ* growth, detachment, and death during the MICP treatment procedure (Wang et al. 2019a), significantly affects crystal number and characteristics (Wang et al. 2021). In this study, we deploy a similar approach to further examine the role of temperature. A series of batch tests were conducted to study the effects of temperature on bacterial population and activity. In addition, microfluidic chip experiments were conducted to observe the MICP processes at the pore scale at different temperatures between 4°C to 50°C. The implications of the findings in terms of resulting engineering properties and treatment protocols for subsurface applications of MICP are also discussed.

## MATERIALS AND METHODS

### Bacterial suspension

*Sporosarcina pasteurii* (DSM 33) was used in the experiments described in this study. Bacterial suspension was prepared using a freeze-dried stock, which was activated according to the supplier's guidelines (DSM). After activation, glycerol stocks of the bacteria were prepared by adding 225 µl of 80 % glycerol (autoclaved) to 1 ml of overnight liquid culture in cryogenic vials, after which the liquid culture was immediately frozen at -80°C (Wang et al., 2019b). Once defrosted, cells from the glycerol stock were grown in ATCC 1376 $NH_4$-YE agar medium (20 g/L yeast extract, 10 g/L ammonium sulphate, 20 g/L agar, and 0.13 M Tris base) for 48 hours at 30°C. Subsequently, several colonies on the agar plate were transferred to a $NH_4$-YE liquid medium containing the same components without agar and cultivated in a shaking incubator for 24 hours at 30°C and a shaking rate of 200 rpm to obtain a bacterial suspension with an optical density measured at a wavelength of 600 nm ($OD_{600}$) of around 3.0. The bacterial suspension with $OD_{600}$ of 1.0 that was used in the experiments was diluted from this bacterial suspension using $NH_4$-YE liquid medium. Based on the results obtained by Wang et al. (2021), the correlation between bacterial number per unit volume, denoted as $C_B$, and bacterial optical density $OD_{600}$ is:

$$C_B = \text{Bacterial density of cells (cells/ml)} = OD_{600} \times 4 \times 10^8 \qquad (3)$$

### Batch test and bacterial activity measurement

Bacteria were cultivated in $NH_4$-YE liquid following the same procedure described in the previous section, to $OD_{600}$ of about 1.0. The bacterial suspension was subsequently divided equally into 12 bottles, with each of the bottle containing about 100 ml of bacterial suspension. Three bottles of bacterial suspension were incubated without shaking at a temperature of either 4, 20, 35 or 50°C. To measure the bacterial activity, 2 ml of bacterial suspension were removed from each of the bottles after every 12 to 24 hours to measure $OD_{600}$. After measuring $OD_{600}$, 1 ml from this 2 ml was mixed with 9 ml of 1.1 mol/L urea for 5 mins to measure the conductivity change and to obtain the ureolysis rate (Wiffen 2007). The conductivity of the mixed content was assessed using a conductivity meter (FiveGo, Mettler-Toledo, Beaumont Leys, Leicester, UK) immediately after the mixing and 5 min after mixing. The ureolysis rate was calculated using Eq. (4) (Whiffin et al. 2007). Measurements were performed in triplicate for each of the different media tested, with data presented as mean ± standard error.

$$U(mM/h) = \frac{\Delta Conductivity(\mu S/cm)}{\Delta t(min)} \times (10^{-3} \times 11.11)(mM/(\mu S/cm)) \times 60(min/h) \qquad (4)$$

The specific urease activity of bacteria was calculated using Eq. (5)

$$U'(mM/h) = \frac{U(mM/h)}{OD_{600}} (mM/h/OD_{600}) \qquad (5)$$

### Cementation solution

The cementation solution for MICP treatment was created using calcium chloride ($CaCl_2$), urea ($CO(NH_2)_2$), and a nutrient broth dissolved in deionized water. The urea and calcium chloride served as important ingredients for promoting calcite precipitation, and the nutrient broth served as an energy source for bacterial activity. The concentrations of $CaCl_2$, urea, and nutrient broth were 0.5 M, 0.75 M and 3 g/L, respectively (Wang et al., 2019b).

### Microfluidic chip experiments and MICP treatment

Microfluidic chip experiments were performed to observe the MICP process in a porous medium under conditions that are affected by flow rate and injection methods (Wang et al., 2019a). The microfluidic chips were designed to create a two-dimensional model of a realistic soil matrix porous structure based on a cross-sectional image of Ottawa 50-70 soil and were fabricated using standard photolithograph techniques. The details of the device and its fabrication process can be found in Wang et al. (2017) and Wang et al. (2019a). Because the microfluidic chip is transparent, the MICP process inside the microfluidic chip can be observed under a microscope. The schematic of the microfluidic chip experiments is shown in Figure 1. Except during bacterial injection (Figure 1a), cementation solution injection (Figure 1a) and imaging (Figure 1c), all samples were kept at the assigned temperatures (Figure 1b).

In total 6 tests were conducted (Table 1). Tests 1-4 were conducted to compare the effects of temperature on MICP. Test 5 was conducted to be compared with Test 4 for investigating the effects of settling time and injection interval of cementation solution on MICP. Test 6 was used to compare with Test 5 for investigating the effects of injection number of bacterial suspension on MICP. In each of the test, after microfluidic chips were made and saturated with deionized water, one pore volume of bacterial suspension with an $OD_{600}$ of 1.0 was injected into each of the microfluidic chip samples at a flow rate of 0.5 ml/h (Darcy velocity of $4.6 \times 10^{-4}$ m/s) for two minutes at room temperature (20°C). The samples were then incubated at different temperature conditions for bacterial settling (Table 1). After settling, in Tests 1 to 5, ten staged injections of cementation solution at a flow rate of 0.05 ml/h (Darcy velocity of $4.6 \times 10^{-5}$ m/s) were conducted. The interval between injections of cementation solution for Tests 1-4 was 48 hours, whereas that for Test 5 was 6~18 h. In Test 6, bacterial injection was performed before every two injections of cementation solution and five injections were performed in total.

**Table 1** MICP treatment parameters for the microfluidic chip experiment

| Test No. | Temperature (°C) | Bacterial injection number | Bacterial settling time (h) | Cementation injection number | Injection interval of cementation solution (h) |
|---|---|---|---|---|---|
| 1 | 4 | 1 | 18 | 10 | 48 |
| 2 | 20 | 1 | 18 | 10 | 48 |
| 3 | 35 | 1 | 18 | 10 | 48 |
| 4 | 50 | 1 | 18 | 10 | 48 |
| 5 | 50 | 1 | 2 | 10 | 6-18 |
| 6 | 50 | 5 | 2 | 10 | 6-18 |

After the MICP treatment procedures were completed, images of samples were taken by an optical microscope (Zeiss Axio Observer Z1) to visualize the micro-scale MICP process (Figure 1c). The microscope was equipped with a digital video camera (Zeiss Axio Observer Z1) connected to a computer. The bacterial cell numbers in the image of one central pore of the microfluidic chips taken after bacterial injection, after bacterial settling, and after the injections of cementation solution were counted to quantify the effects of temperature on bacterial growth, attachment, and detachment.

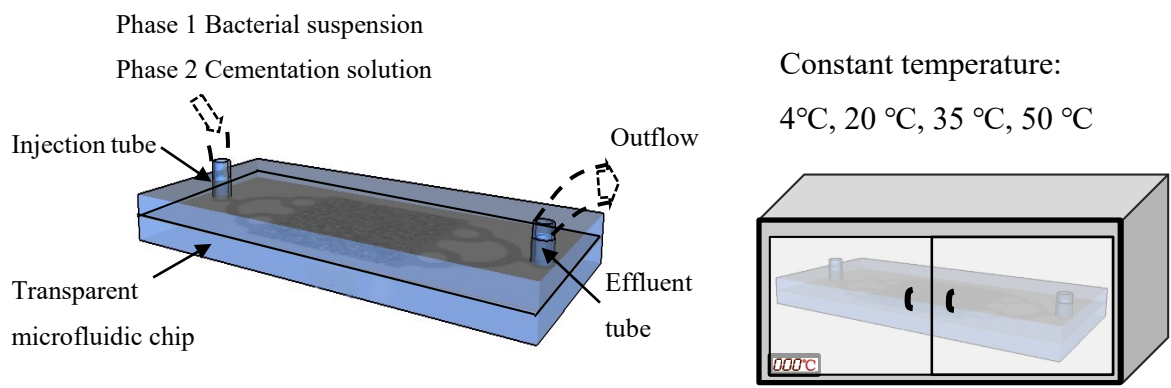

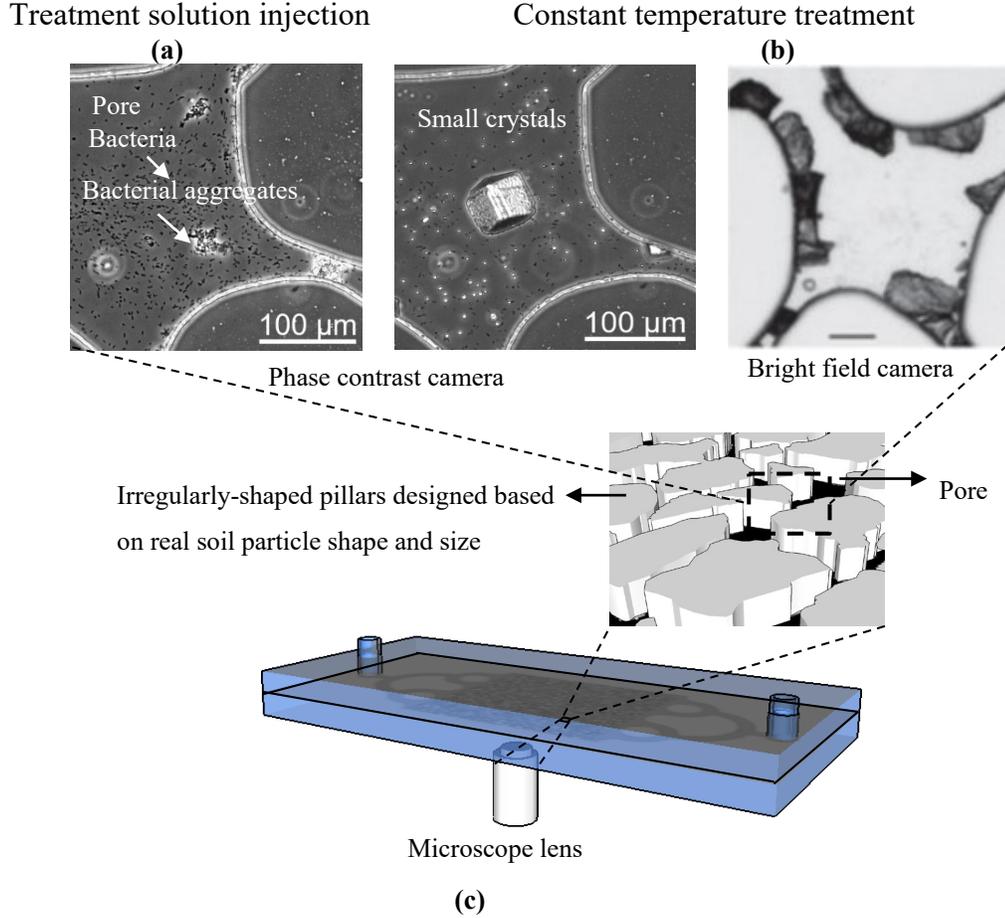

**Figure 1** Schematic of microscale microfluidic chip experiments, (a) bacterial and cementation solution injection; (b) incubation; (c) imaging

**Image quantification**

Two types of CaCO$_3$ crystals were mostly reported in previous studies, namely spherical and rhombohedral crystals (Wang et al. 2019b, 2021). The observed CaCO$_3$ crystals that grow in the microfluidic chip channels are usually semi-spherical and semi-rhombohedral crystals (Wang et al. 2019b, 2021). To quantify the size of semi-spherical crystals, the diameters of the crystals in the 2-D images were measured by ZEN software (Zeiss) and the volumes of the half spheres were calculated. The schematics of the crystal volume calculation is shown in Figure 2. Because the depth of the microfluidic channels is 50 μm, when the diameter of the sphere is smaller than 100 μm, the crystal volume is calculated as the volume of a half sphere:

$$V = \frac{1}{2} \times \frac{4}{3}\pi \left(\frac{D}{2}\right)^3 = \frac{1}{12}\pi D^3 \tag{6a}$$

where D is the diameter of the half-sphere. When the diameter of the sphere is larger than 100 μm, the crystal volume is the half sphere minus volume of the cap of height, which is calculated as:

$$V = \frac{1}{12}\pi D^3 - \frac{\pi}{3}(50+D)\left(\frac{D}{2} - 50\right)^2 \tag{6b}$$

To quantify the size of semi-rhombohedral crystals, the size of the bottom square of the crystals in the 2-D images was measured by ZEN software (Zeiss), and the height of the semi-rhombohedral crystals (denoted as a) is assumed to be the same as the length of the bottom square of the crystals. Therefore, when the height is smaller than 50 μm, the crystal volume can be calculated as:

$$V = \frac{1}{2}a^3 \tag{7a}$$

When the height of the crystal is larger than 50 μm, the crystal volume can be calculated as:

$$V = \frac{1}{2}a^3 - \frac{1}{2}(a-50)^3 \tag{7b}$$

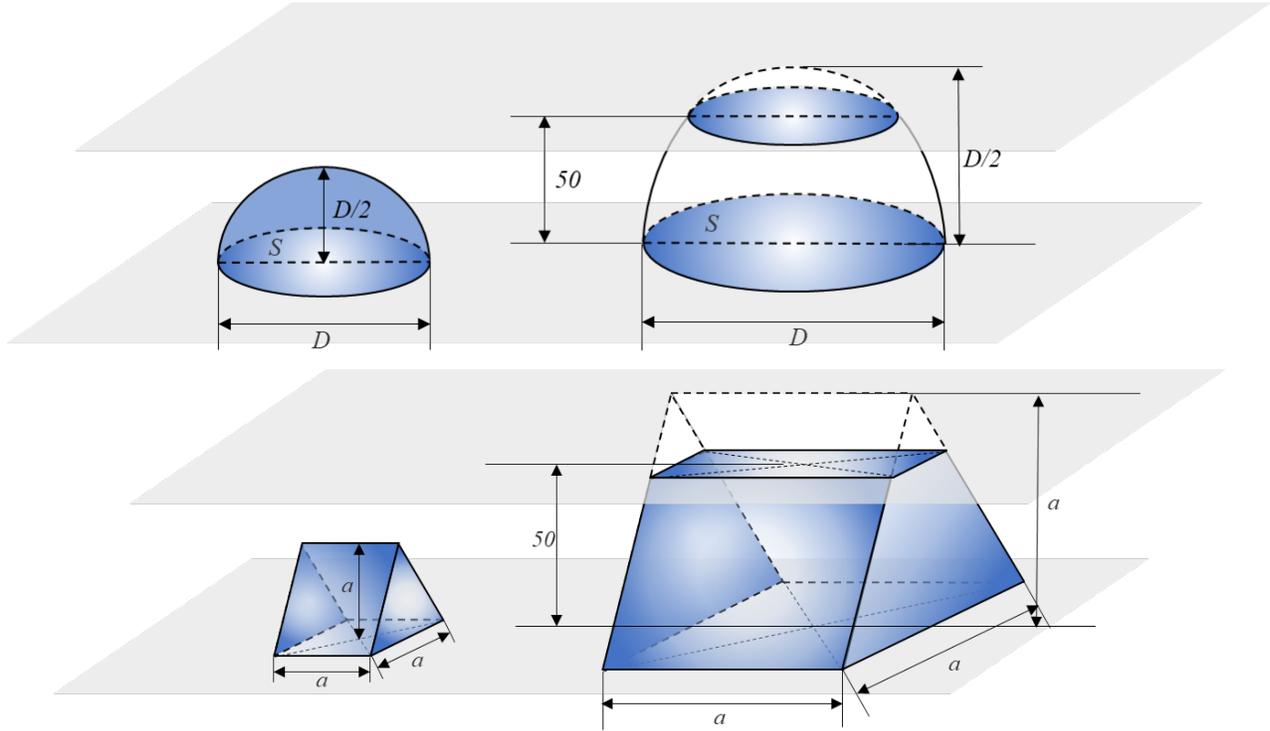

**Figure 2 Schematic of spherical and rhombohedral crystals**

The number of crystals present in selected areas of the images were counted and crystal densities were calculated by dividing the number of crystals by the volume of the pores that contain the crystals.

Chemical transformation efficiency (CTE) after the $n_{th}$ injection of cementation solution indicates the actual mass or volume of calcium carbonate obtained after MICP treatment relative to the theoretical calculated mass or volume of calcium carbonate obtained if all of the calcium ions injected transformed into calcium carbonate (Al Qabany et al. 2012). The total volume of $CaCO_3$ (assuming all crystals are calcite) is denoted as $V_{c100\%}$, the pore volume is denoted as $V_v$, and the ratio of $V_{c100\%}$ relative to $V_v$ can be calculated by:

$$\left(\frac{V_{c100\%}}{V_v}\right)_n = \frac{\dfrac{0.5\ mol/L \times n \times 100\ g/mol \times PV}{2.71\ g/cm^3} \times 0.001\ L/cm^3}{PV} \times 100\% \tag{8a}$$

After the $n_{th}$ injection, the total volume of $CaCO_3$ crystals as a fraction of the pore volume $V_v$ is $(V_c/V_v)_n$. The chemical transformation efficiency after the $n_{th}$ injections of cementation solution is:

$$(CTE)_n = \frac{\left(\dfrac{V_c}{V_v}\right)_n}{\left(\dfrac{V_{c100\%}}{V_v}\right)_n} \times 100\% = \left(\dfrac{V_c}{V_{c100\%}}\right)_n \times 100\% \qquad (8b)$$

The individual chemical transform efficiency can be computed to determine the chemical transform efficiency after each of the injections of cementation solution. After each injection of cementaion solution, if all of the injected $Ca^{2+}$ ions fully transform into $CaCO_3$, then the total volume of calcium carbonate crystals (assuming all crystals are calcite) produced by one injection $V_{c100\%}$ relative to the pore volume $V_v$ is:

$$\frac{V_{c100\%}}{V_v}{'} = \frac{\dfrac{0.5 mol/L \times 100 g/mol \times PV}{2.71 g/cm^3} \times 0.001\, L/cm^3}{PV} \times 100\% \qquad (9a)$$

As a result, the chemical transformation efficiency following an individual injection after the $n_{th}$ injection, indicated by the volume of crystals measured after the completion of the $n_{th}$ injection as a fraction of the pore volume $V_v$ minus the volume of crystals measured after the completion of the $(n-1)_{th}$ injection as a fraction of the pore volume $V_v$:

$$(ICTE)_n = \frac{\left(\dfrac{V_c}{V_v}\right)_n - \left(\dfrac{V_c}{V_v}\right)_{n-1}}{\dfrac{V_{c100\%}}{V_v}{'}} \times 100\% \qquad (9b)$$

## RESULTS AND DISCUSSION
### Effects of temperature on bacterial density and activity changes

The change in bacterial $OD_{600}$, ureolysis rate and specific urease activity at 4, 20, 35 and 50 °C (Tests 1-4) are shown in Figure 3 a, b and c, respectively. In all four temperature conditions, the $OD_{600}$ values reduced with time over the 12 days, with the reduction rate being higher at higher temperatures (Figure 3a). The reduction rate of $OD_{600}$ is about 0.01, 0.04, 0.2 and 0.4 per day when temperature was 4, 20, 35 and 50°C, respectively (Figure 3a). Temperature also affects the changes of ureolysis rate (Figure 3b). At 4 and 20°C (Tests 1 and 2, respectively), the ureolysis rate reduced from 40 mM/h to about 25 and 30 mM/h, respectively, after about 2 days, and remained almost constant until the end of the 12-day period (Figure 3b); at 35°C (Test 3), the ureolysis rate increased to about 60 mM/h within approximately two days, and then started decreasing with time until 0 by about 12 days; at 50°C (Test 4), the ureolysis rate reduced significantly with time to zero within just one day (Figure 3b). To remove the effects of bacterial density on ureolysis rate, bacterial specific urease activity is plotted against time in Figure 3c. At 4°C, bacterial specific urease activity remained almost constant at about 34 mM/h/$OD_{600}$; at 20°C it increased slowly from about 35 mM/h/$OD_{600}$ to 50 mM/h/$OD_{600}$ within twelve days; at 35°C, it increased linearly from about 35 mM/h/$OD_{600}$ to about 250 mM/h/$OD_{600}$ within 4.75 days, and then reduced linearly to zero within 6.75 days; at 50°C, it decreased linearly to zero within just one day (Figure 3c).

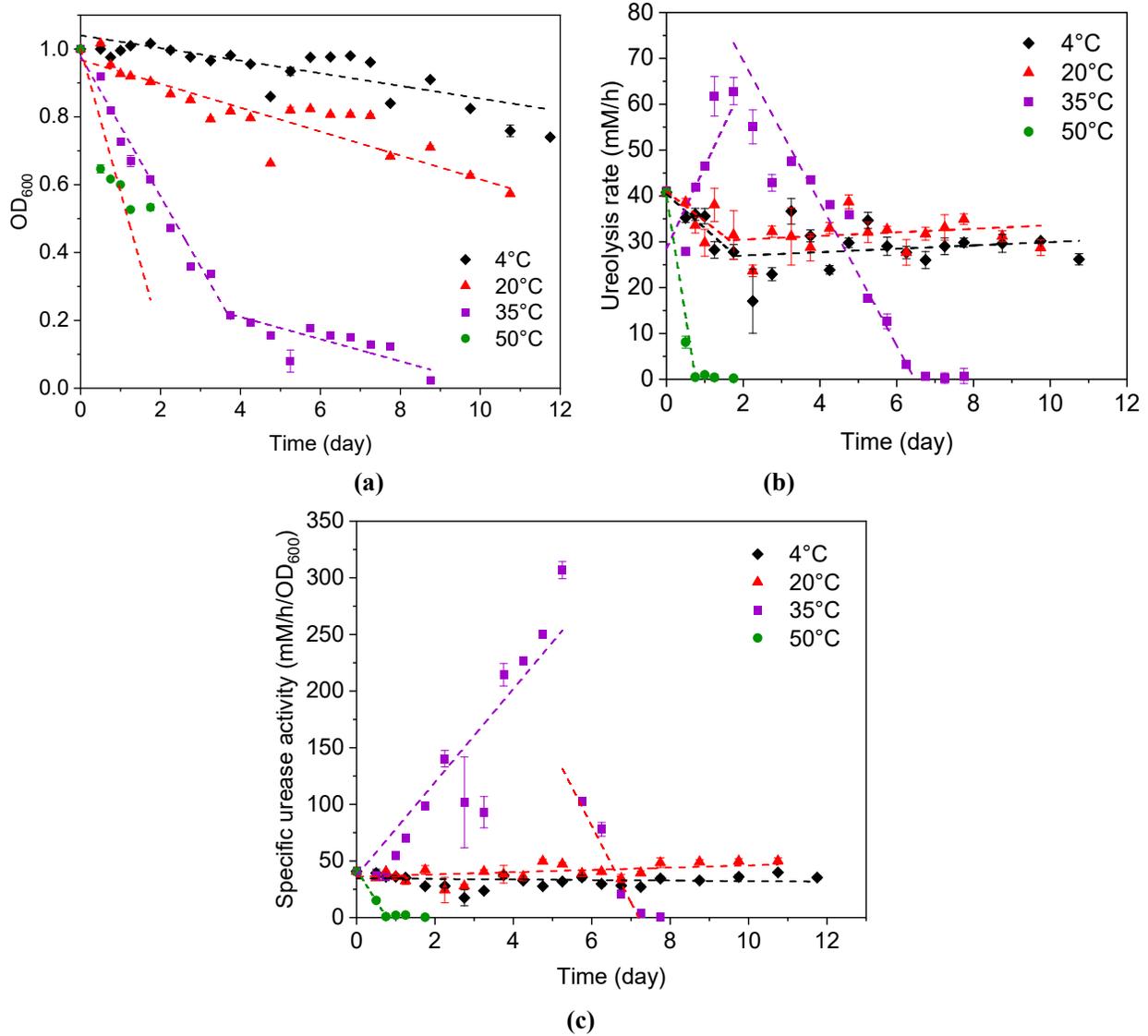

**Figure 3** Results of batch tests (a) optical density of bacterial suspension v.s. time; (b) ureolysis rate of bacterial suspension v.s. time; (c) specific urease activity v.s. time;

Ureolysis rate is mainly affected by bacterial density $C_B$ (cells/l), bacterial urease producing rate $C_u$ (g/cells), and pure urease activity $A_u$ (mol/l/h/g). These three parameters are all functions of temperature T and time t. The ureolysis rate can be expressed as:

$$U(T, t) = C_B(T, t) \times C_u(T, t) \times A_u(T, t) \qquad 10$$

Compared with Equations. 3-5, bacterial specific urease activity is a function of the rate of urease production and urease activity.

$$U'(T, t) = C_u(T, t) \times A_u(T, t) \qquad 11$$

The ureolysis rate can be expressed as:

$$U(T, t) = C_B(T, t) \times U'(T, t) \qquad 12$$

Bachmeier et al. (2002) investigated the effects of temperature on pure free enzyme activity and found that urease activity decreases with time at different temperatures, with higher temperatures reducing the activity more quickly than lower temperatures. The percentage of remaining urease activity of the free enzyme after 1 day in their experiments was only about 90%, 60% and 10% of the original activity at temperatures of 4, 30 and 60 °C, respectively. Fathima and Jayalakshmi (2012) observed that urease had maximum activity at 35°C. Khan et al. (2019) studied the urease production ability of bacterial strains at different temperatures and found that urease production rates increase with temperature until the temperature reached 35°C, after which urease activity started decreasing until 50°C.

In this study, the reduction in ureolysis rate at 4°C (Test 1, Figure 3b) was mainly caused by bacterial density $C_B$ which reduced with time (Figure 3a), but was unaffected by bacterial specific urease activity U' which remained constant over the 12-day period (Figure 3c). The ureolysis rate was higher at 20°C (Test 2) than at 4°C (Figure 3b). This is because, although the bacterial density decreased by a greater extent at 20°C than at 4°C (Figure 3a), the increase in U' at 20°C compensated for the decrease in bacterial activity. According to the study of Khan et al. (2019), the increase in U' might be due to the higher urease production rate at 20°C than at 4°C. At 35°C (Test 3), even though the optical density of bacterial suspension reduced to a greater extent with time than at 20°C or 4°C (Figure 3a), ureolysis rate still increased with time during the first two days (Figure 3b). This is because an increase in bacterial specific urease activity compensated for a decrease in bacterial density (Figure 3a and c). However, after two days, urease activity at 35°C reduced with time and this is because the increase in bacterial specific urease activity could not compensate for the decrease in bacterial density (Figure 3a and c). At 50°C (Test 4), the reduction in bacterial urease activity with time (Figure 3b) was due to the reduction in both bacterial density (Figure 3a) and bacterial specific urease activity (Figure 3c) with time.

**Effects of temperature on bacterial growth and attachment**

The effects of temperature on bacterial growth and attachment were investigated using microfluidic chip experiments. The microscope images are shown in Figure 4a and the quantification results are shown in Figure 4b. Although the bacterial density after injection was similar (see the first row of Figure 4a and the white columns of Figure 4b), the bacterial density after the injections of cementation solution varied (see the third row of Figure 4a and the dark grey columns of Figure 4b). This is due to both the difference in amount of *in-situ* bacterial growth (shown by the differences between the first row and the second row of Figure 4a and between the light grey and dark grey columns in Figure 4b), and the difference in bacterial detachment by the injection of cementation solution (shown by the differences between the first row and the second row of Figure 4a and between the light grey and dark grey columns in Figure 4b). Optimal growth of bacteria was observed when the temperature was 20°C (Test 2) and optimal attachment of bacterial was observed when the temperature was 35°C (Test 3). After the injections of cementation solution, the highest bacterial density was obtained when the temperature was 35°C, while the lowest bacterial density was obtained when temperature was 4°C. The bacterial densities were about 0.3, 3, 3.5, and 1.1 times of their initial values for the temperature conditions of 4, 20, 35 and 50℃, respectively. Therefore, when considering the effects of bacterial growth, the bacterial density $C_B$ in Equation 9 is a function of temperature, time, nutrient /oxygen condition, flow rate, solid surfaces conditions etc.

**Temperature-dependent crystal growth procedure after the 1$^{st}$ and 2$^{nd}$ injections**

Microscopic images of one pore in the central region of the microfluidic chips at 3, 6, 24, and 48 hours after the first and second injections of cementation solution are shown in Figure 5 for the four temperature conditions (4, 20, 35 and 50°C, Tests 1-4). Chemical transformation efficiencies, crystal number and crystal precipitation rates of crystals shown in Figure 5 were quantified, and the results are shown in Figure 6 a, b and c.

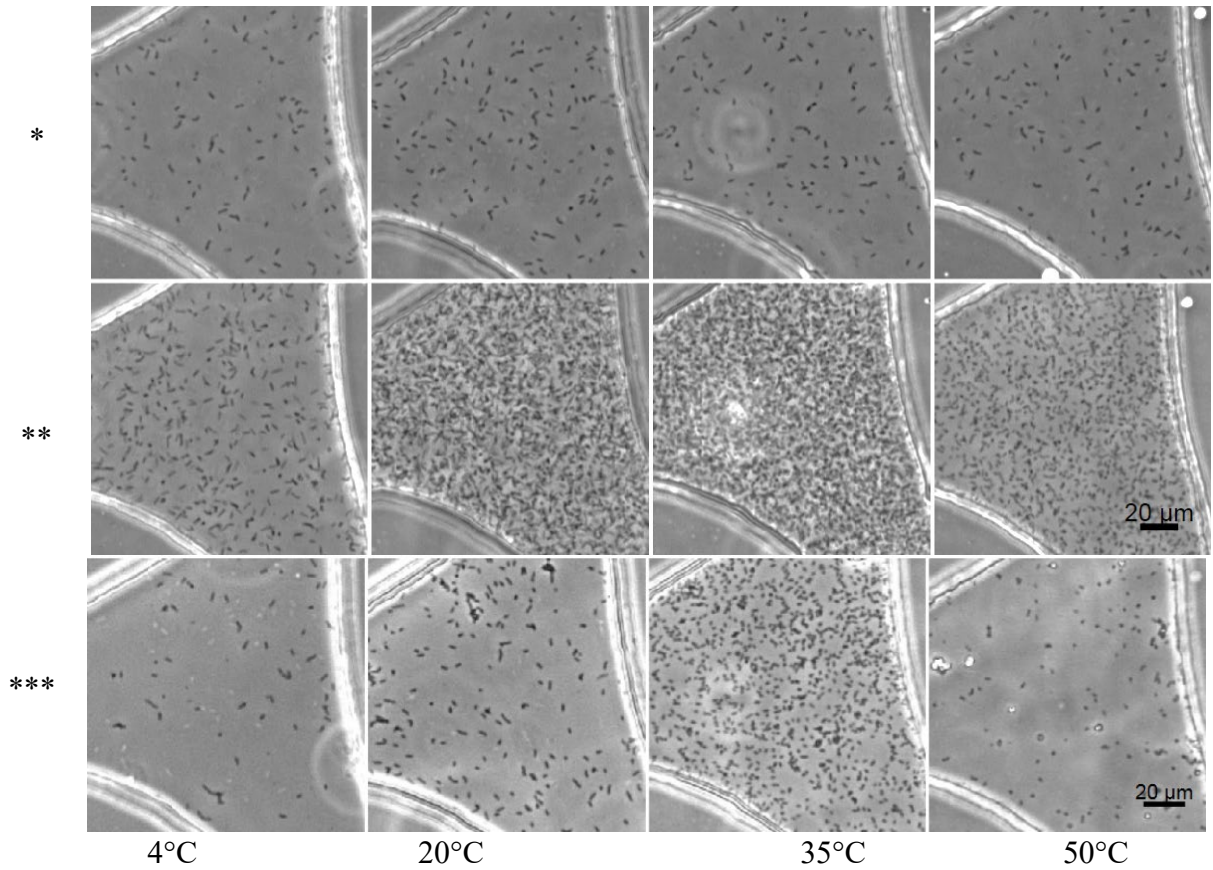

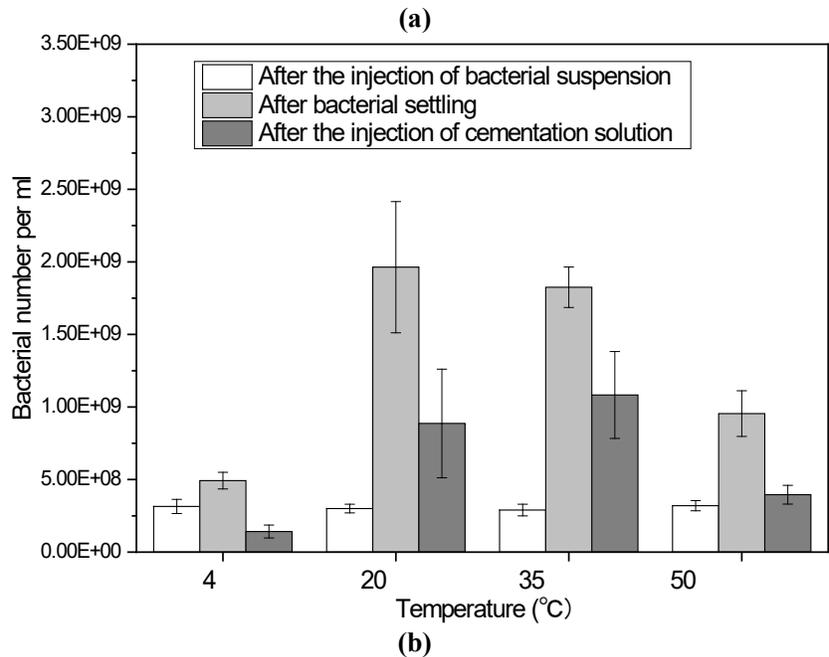

**Figure 4 (a) Microscope images of a middle pore of the microfluidic chips taken after bacterial injection (\*), after bacterial settling (\*\*) and after cementation solution injection (\*\*\*); (b) the quantification of the bacterial densities in images shown in (a)**

It can be seen from Figure 5 and Figure 6a that crystals grow more slowly at 4°C compared to the other three temperature cases. For the temperature cases of 20°C, 35°C, and 50°C, after the first injection of cementation solution, crystals completed precipitation within 6 hours, whereas for the 4°C case crystals did not complete their precipitation by 48 hours. In addition, the crystal growth rates after the first injection of cementation solution were also different to those after the second injection, especially at 50°C, where the time for the crystals to complete the precipitation after the first and second injections of cementation solution were 6 and 48 hours, respectively. Moreover, the growth rate at 20°C is consistent with the results of Wang et al. (2019b), wherein for a cementation solution of concentration 0.25 M, the time for crystals to fully precipitate is about 3 hours. In the present study, the concentration of cementation solution is doubled and hence the time required to precipitate doubled as well.

It can be seen from Figure 5 and Figure 6b that crystal number increases due to the growth of new crystals and decreases due to the dissolution of unstable crystals. At 4 and 20°C, crystal number mainly increases after the first two injections of cementation solution. However, at 35°C and 50°C, a crystal precipitation-dissolution-reprecipitation process was observed, as shown by the white circles in the second rows of Figure 5c and 5d. With the dissolution of unstable crystals, the crystal number also reduced (Figure 5b). In addition, the dissolution mainly occurred between 6 and 24 hours after the first two injections of cementation solution at 35°C, and after the second injections of cementation solution at 50°C (Figure 5c, d and Figure 6b).

The precipitation-dissolution-reprecipitation process of these crystals is consistent with Ostwald's law of crystal ripening; usually, the least dense phase is formed first and transforms to the next dense phase until finally the densest (which is usually also the most stable phase) is formed (reviewed by Cöelfen and Antonietti, 2008). The shapes and stabilities of the spherical and rhombohedral $CaCO_3$ crystals are consistent with those of vaterite and calcite, respectively, which is consistent with the previous studies of Wang et al. (2019b, 2021). Calcite is the most stable form of $CaCO_3$ crystals. In addition, the dissolved crystals have a 'memory-like' effect; that is, when a new batch of cementation solution is injected, the crystals tend to reprecipitate where the crystals used to be located after the previous injection before they dissolved (see the changes of the crystal indicated by the yellow arrows in Figure 5c). This is probably because the crystals did not fully dissolve, with some crystal lattice still remains on the channel surface of the porous medium. Crystals grow on existing lattices because less energy is required than growing on a new surface due to the extra energy required for nucleation (Kralj et al. 1997).

At 4°C, the chemical transformation efficiencies were around 64% by 48 hours after the first injection of cementation solution (Figure 6a). Based on this rate an injection interval of approximately 75 hours is required to achieve a high chemical transform efficiency (about 100%). In addition, based on the chemical transform efficiency at 3 hours in the 20°C (Test 2), 35°C (Test 3), or 50°C (Test 4) cases (Figure 6a), the time required for 0.5 M of cementation solution to complete precipitation after the first injection of cementation solution is 6, 3.3, or 3.75 hours, respectively. Considering the difference in bacterial density after the injections of cementation solution in these four cases (shown in Figure 3b) and given that the bacterial suspension has $4 \times 10^8$ cells per ml when $OD_{600}$ is 1.0 based on results of Wang et al. (2021), the calculated precipitation rates in these four conditions are 0.018, 0.038, 0.056, and 0.135 M/h/OD, respectively (Figure 6c). As the temperature increases, crystal growth rate increases exponentially (Figure 6c).

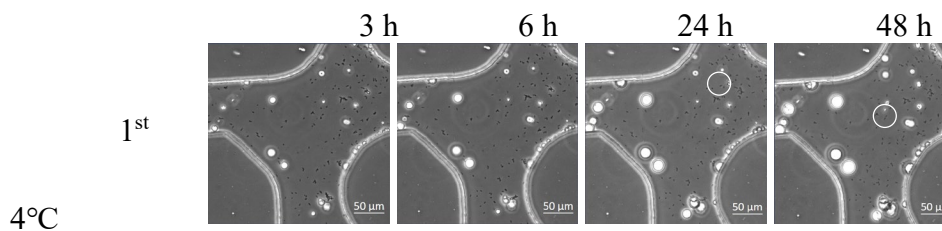

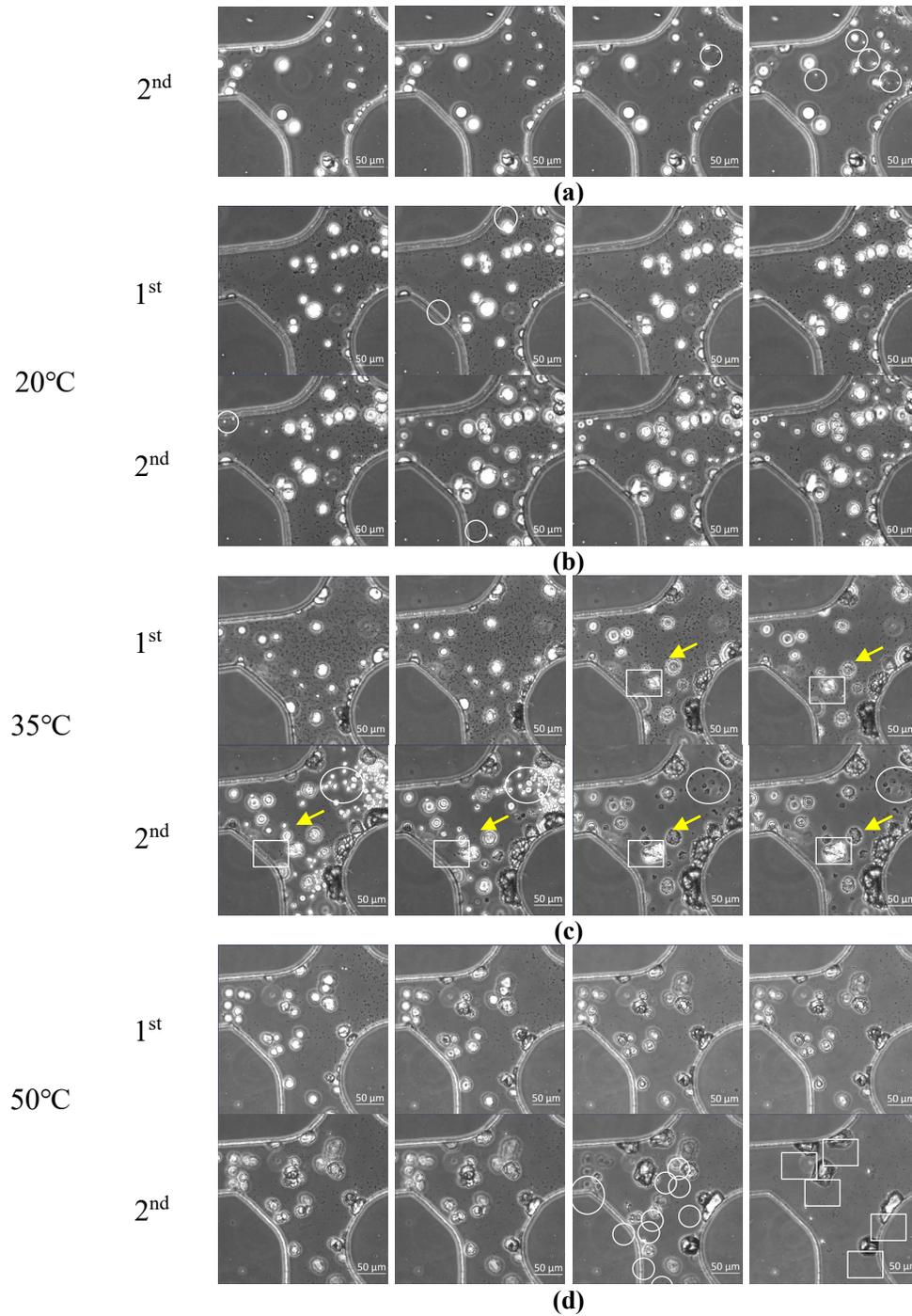

**Figure 5** Sequential microscope images taken on one pore at the central region of the microfluidic chips placed at 4, 20, 35 and 50°C at 3, 6, 24 and 48 hours after the first and second injection of cementation solution, (a) 4°C, (b) 20°C, (c) 35°C and (d) 50°C.

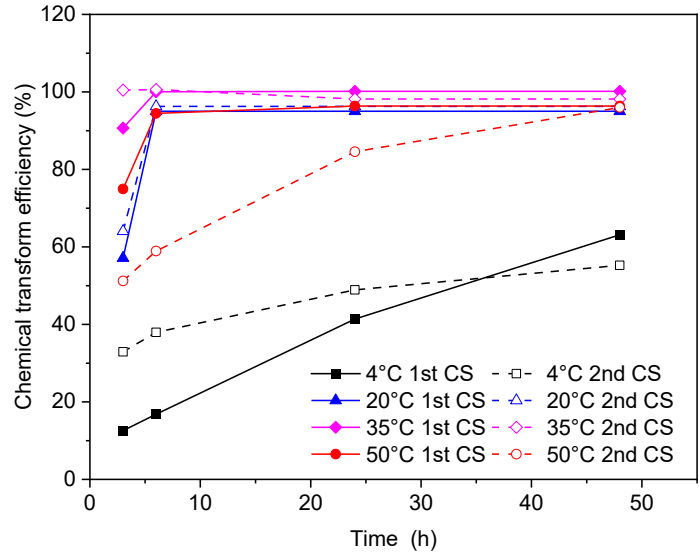

(a)

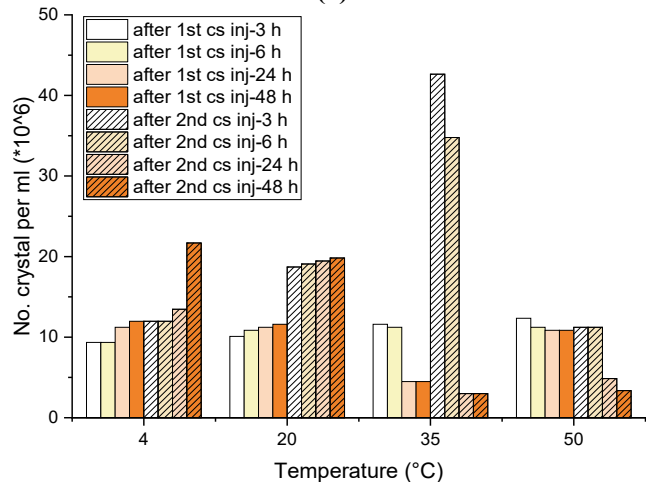

(b)

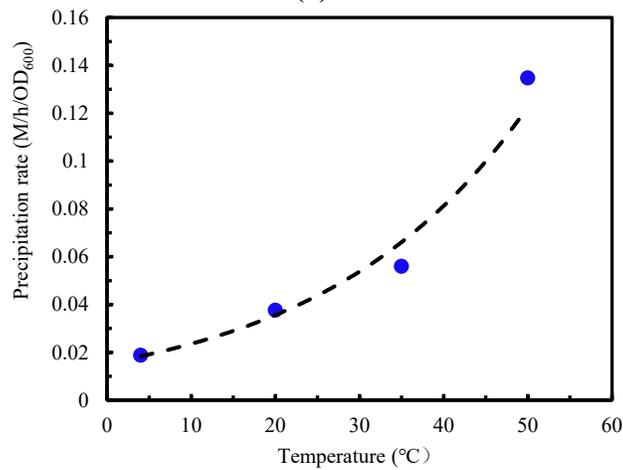

(c)

**Figure 6 Quantification results of images shown in Figure 5. (a) chemical transform efficiency; (b) crystal number; (c) precipitation rate.**

**Temperature-dependent crystal growth procedure after the 2th to 10th injections**

To observe the crystal growth during the entire treatment process, microscope images of one pore located near the center of the microfluidic chip were taken two days after the injections of the second, fourth, sixth, eighth, and tenth injections of cementation solution and are shown in Figure 7. The quantified average crystal volume, crystal number per unit pore volume, ratio of crystal volume to pore volume ($V_{crystals}/V_{pore}$), total chemical transformation efficiency, and individual chemical transformation efficiency are given in Figure 8a-e.

The average crystal volume for the three cases grew steadily as the number of cementation solution injections was increased (Figure 7 and Figure 8a), except for the 50°C case (Test 4) where individual crystals stopped growing after the fourth injection of cementations solution. At 4, 20, 35°C and 50°C (Tests 1-4), the average crystal size increased at rates of 402.4, 1344.3, 1881.8 and 312.5 $\mu m^3$ per injection, respectively. After the completion of the tenth treatment, the average crystal size was the largest when the temperature was 35°C (Test 3), and the smallest when the temperature was 4°C (Test 1, Figure 7 and Figure 8a).

Crystal number can increase due to the precipitation of new crystals and decrease due to the dissolution of unstable crystals (Figures 5 and 7). At 4°C (Test 1), crystal number remained constant (Figure 5), possibly because this low temperature does not provide enough energy for new crystals to nucleate, and because dissolution of crystals does not occur. At 20°C (Test 2), the number of crystals decreased with time, indicating that dissolution of crystals occurred at 20°C, but the dissolution rate was lower than at 35°C (Test 3) and 50°C (Test 4), shown by the much higher crystal number at 20°C than at 35°C or 50°C (Figure 8b). At 35°C, the dissolution procedure completed within the 48 hours after each injection and the growth of new crystals appeared constant after each injection (Figure 7 and Figure 8b). At 50°C, MICP stopped because the rate of urease activity reduced to zero. In addition, the dissolution procedure completed within the 48 hours after each of the injections; therefore, crystal number remained constant (Figure 8b). Even though the crystal number remained the same at both 4°C and 50°C, the number of crystals observed at 4°C was much higher than at high temperature because crystal dissolution occurred at 50°C but not at 4°C (Figure 8b).

The total crystal volume is indicated by the volume ratio of crystal volume to pore volume and is plotted against the measured crystals number in Figure 8c. At 4, 20, and 35°C crystals grew steadily at almost constant rates, indicated by the ratio of total crystal volume to pore volume (Figure 8c), with the highest rate being observed at 20°C, while the lowest rate was observed at 4°C. At 50°C, crystals stopped growing after the fourth injection of cementation solution (Figure 8c), and the growth rates after the first four injections of cementation solution are similar to those at 4°C.

At these four temperatures, the total chemical transform efficiency varied significantly and decreased linearly with time at different rates. The initial chemical transform efficiencies obtained at 35°C and 20°C were about 100% and 90%, respectively. They were much higher than the initial chemical transformation efficiencies at 50°C and 4°C, which were about 60% and 50%, respectively. However, due to the difference in reduction rate of chemical transformation efficiency with time, at 4, 20, 35 and 50°C, the chemical transformation efficiencies after the 10th injection were about 30%, 80%, 70%, and 30%, respectively (Figure 8d). The individual chemical transformation efficiencies decreased at similar rates when the temperatures were 4°C and 20°C (about 4% per injection), but reduced much more quickly when the temperature was 35°C. The efficiency reduced even more quickly when the temperature was 50°C where the transformation efficiency reduced to zero by the 6th injection of cementation solution (Figure 8e).

In summary, temperature has both a direct effect and indirect effect on crystal growth rate. The direct effect occurs when the supersaturation ratio is the same, with an increase in temperature leading to an increase in the CaCO$_3$ precipitation rate. The indirect effect is that temperature affects the bacteria quantity and activity, which in turn affects the rate of carbonate production, and consequently affects the supersaturation condition. For example, the crystal growth rate is higher at 4°C than at 50°C. Although the number of bacteria remaining after the injection of cementation solution is higher at 50°C than at 4°C and CaCO$_3$ crystal growth rate is usually higher at higher temperatures when the supersaturate ratio is the same, the specific bacterial activity at 50°C decreases significantly with time and reduces to zero after a relatively very short period of time. At 4°C, however, the bacterial activity remains constant at a relatively high rate. The same reason is valid for the comparison between the 35°C case and the 20°C case. At 35°C, the bacterial density is higher than that at 20°C, but because at 35°C the bacterial activity increases to peak and then reduces with time, the MICP rate and efficiency are lower than those at 20°C.

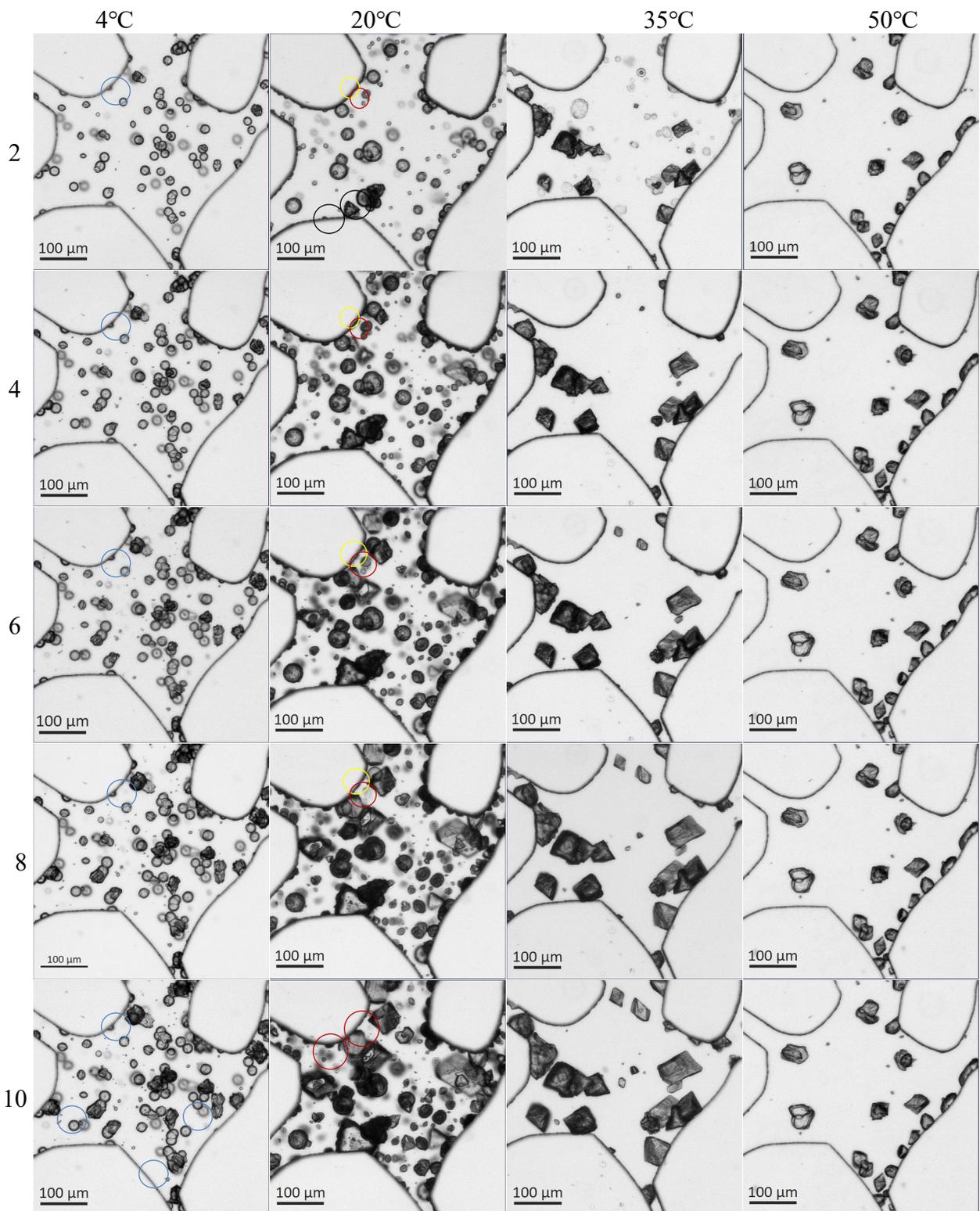

**Figure 7** Microscope images taken on one pore at the central region of the microfluidic chips placed at 4, 20, 35 and 50°C at two days after the 2nd, 4th, 6th, 8th and 10th injections of cementation solution

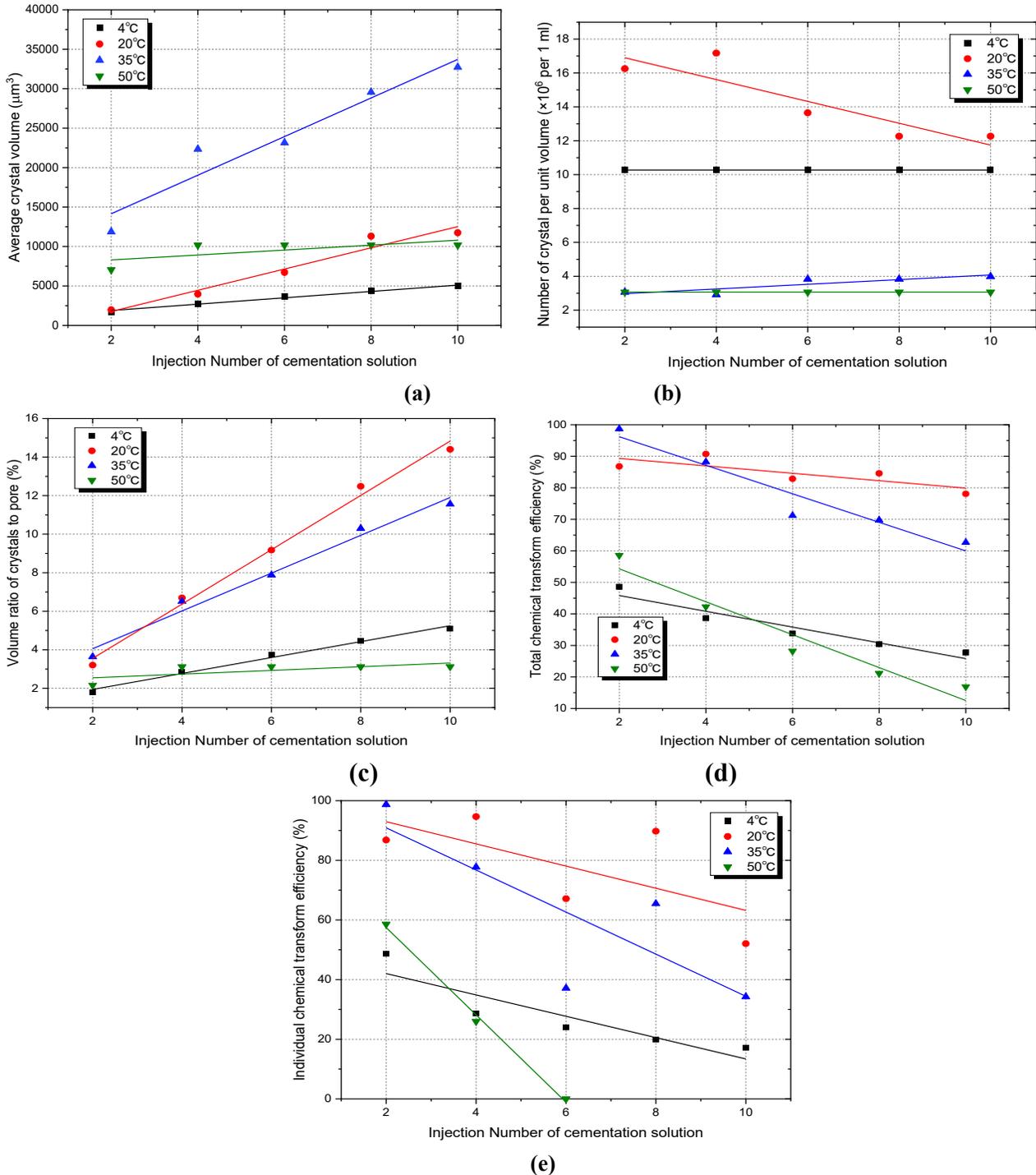

**Figure 8** Quantification results of Figure 7 (a) Average crystal volume v.s. injection number of cementation solution; (b) number of crystals per unit pore volume v.s. injection number of cementation solution; (c) total crystal volume/Pore volume v.s. injection number of cementation solution; (d) chemical transform efficiency v.s. injection number of cementation solution; (e) Individual chemical transform efficiency v.s. injection number of cementation solution

Han et al. (2021) found that CaCO$_3$ content obtained at 50°C is higher than the CaCO$_3$ content obtained at 35°C. This is contrary to the results herein, where after 10 injections of cementation solution the CaCO$_3$ content obtained at 35°C is much higher than the CaCO$_3$ content obtained at 50°C. The reason might be that a 18-hour bacterial settling time was set in this study, over which the bacterial activity decreased significantly at 50°C, but increased at 35°C. Therefore, when the cementing solution was introduced into the system after settling, 50°C resulted in low precipitation and 35°C resulted in high precipitation. However, in the study of Han et al. (2021), there was not a settling time defined. Therefore, bacterial setting time might be an important factor to consider for MICP studies, especially when bacterial activity changes over the settling hours.

**Effects of temperature on crystal dissolution**

To observe the crystal dissolution in detail, the images of the centre pores at 4, 20, 35, and 50°C (Tests 1-4) at two days after the 1$^{st}$, 2$^{nd}$, 4$^{th}$, 6$^{th}$, 8$^{th}$, and 10$^{th}$ injections of cementation solution are shown in Figure 9. The crystals that used to exist at some point but dissolved at later stages are circled in red. Dissolution occurred at temperatures of 20, 35, and 50°C, but not at 4°C. At 20°C, some of the crystals dissolved between the 4$^{th}$ and 6$^{th}$ injections (shown by the red circles in the 4$^{th}$ injection image) and between the 6$^{th}$ and 10$^{th}$ injection (shown by blue arrows where the crystal colour became lighter indicating crystal dissolution). At 35°C and 50°C, the crystals dissolved between the first and the second injections, but not afterwards. This result is the same as shown in Figure 8b. After the 2$^{nd}$, 4$^{th}$, 6$^{th}$, 8$^{th}$, and 10$^{th}$ injections of cementation solution, the number of crystals decreased gradually at 20°C, whereas the number of crystals remained almost constant at 35°C and 50°C.

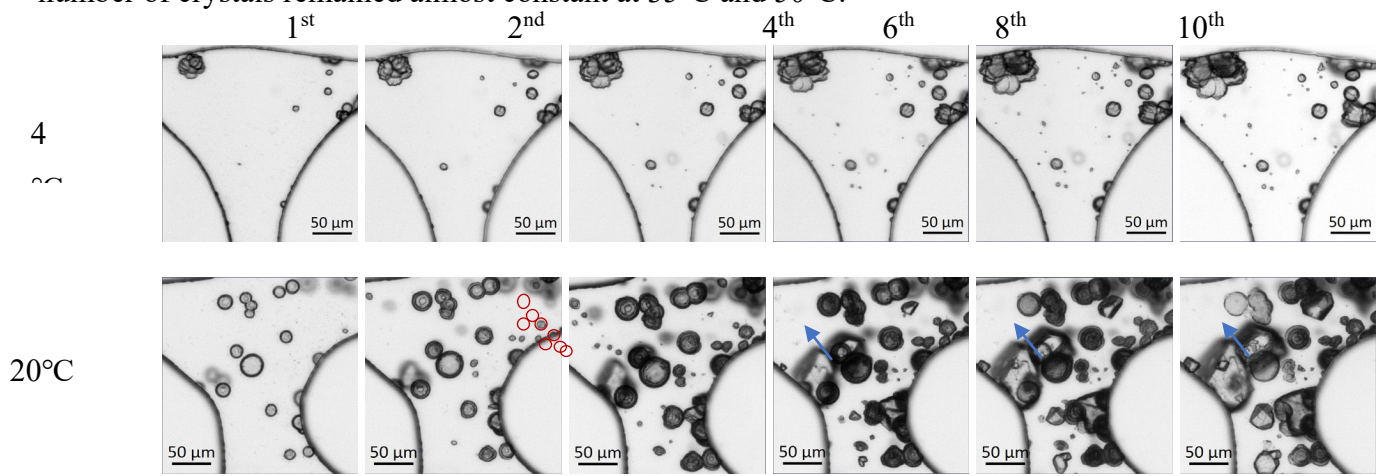

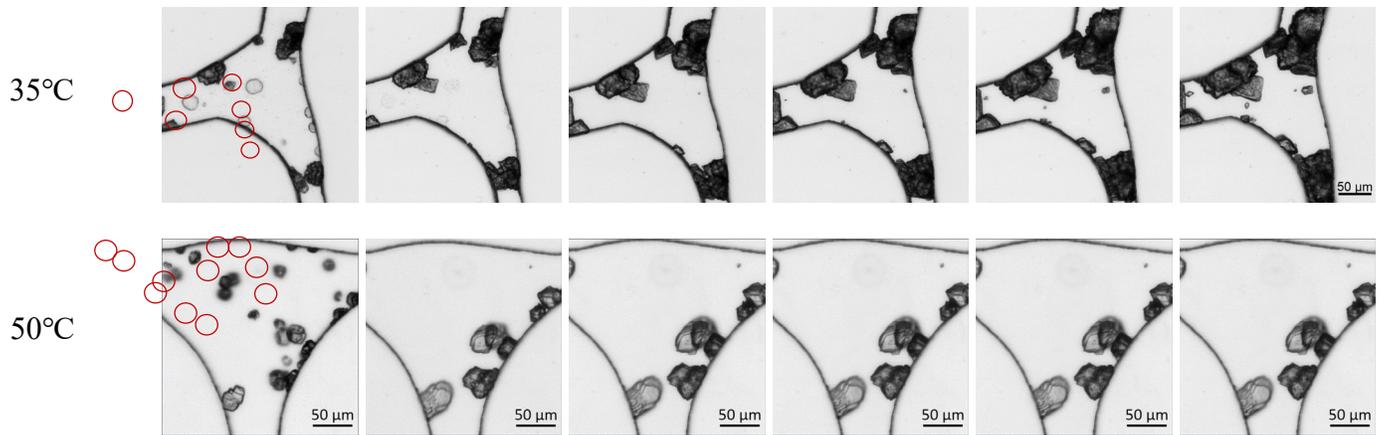

**Figure 9 Microscope images at two days after the 1st, 2nd, 4th, 6th, 8th, and 10th injections of cementation solution**

To observe a larger area, the images of a 1.5 mm by 1.5 mm square area taken at the middle of the microfluidic chip at 2 days after the first and tenth injection of cementation solution are shown in Figure 10. Crystal dissolution occurred at 20°C, 35°C and 50°C (Tests 2-4), but not at 4°C (Test 1). In addition, a rise in temperature promoted crystal dissolution as shown by the increase in numbers of red circles from 4°C to 50°C. This trend is consistent with the results of Kralj et al. (1997) where a pure chemical precipitation of $CaCO_3$ and transformation from vaterite to calcite at, 25, 35, 40, and 45°C were studied.

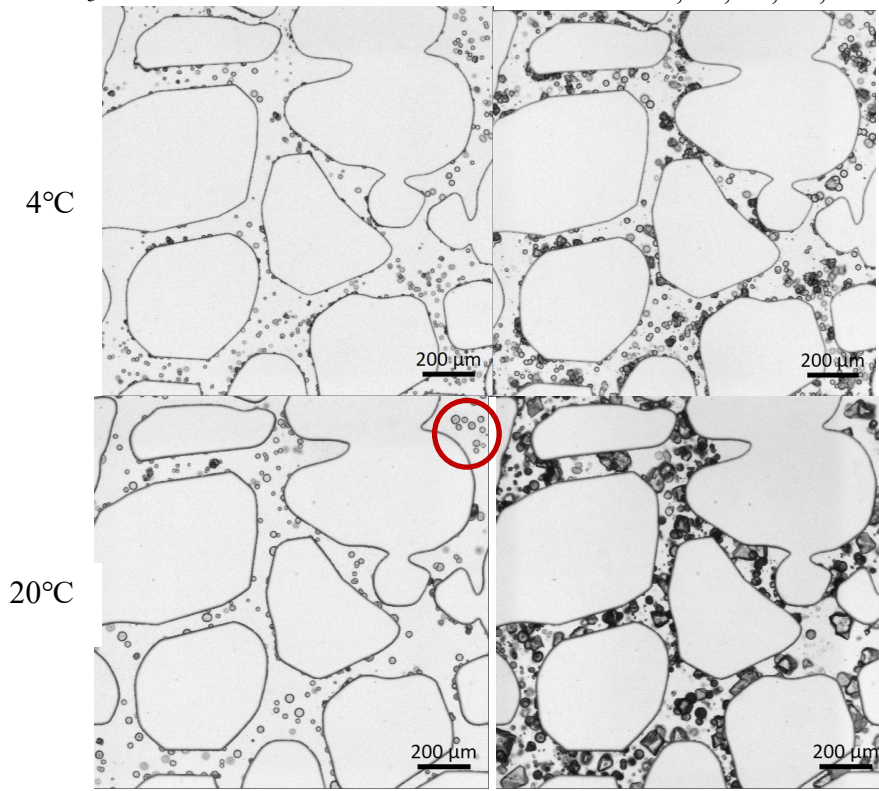

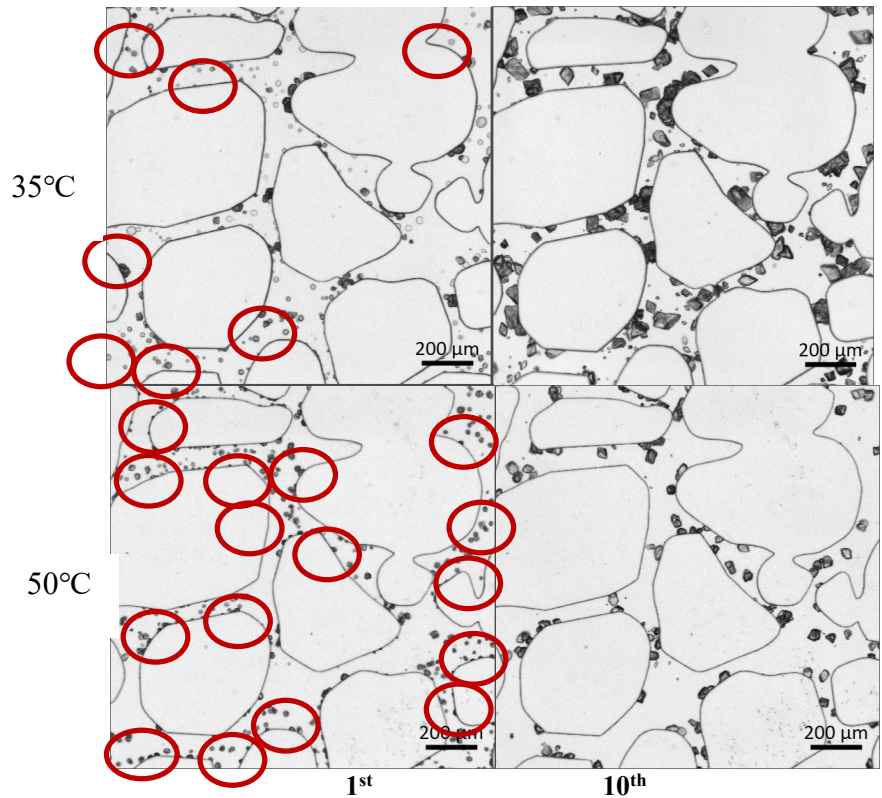

**Figure 10 1500 μm by 1500 μm sized microscope images taken at the middle of the microfluidic chips at 2 days after the 1st and 10th injections of cementation solution**

**Effects of temperature on crystal characteristics**

To observe crystal characteristics, microscope images taken at the centre of the microfluidic chips at the completion of the MICP treatments at these four temperature conditions are shown in Figure 11. The magnified images of four pores presented in Figure 11 are shown in Figure 12 to examine the details. Crystals were mainly spherical at 4°C (Test 1), appeared as a mixture of spherical and rhombohedron crystals at 20°C (Test 2), while rhombohedral calcite crystals were observed at 35°C (Test 3) and 50°C (Test 4). The ratio of the number of rhombohedral crystals to the total number of crystals was 20%, 35%, 100%, and 100% at 4°C, 20°C, 35°C and 50°C, respectively. Rhombohedral and spherical crystal shapes are consistent with the morphology of calcite and vaterite, respectively. This suggests that high temperature conditions favoured the formation of calcite, whereas low temperature conditions favoured the formation of vaterite.

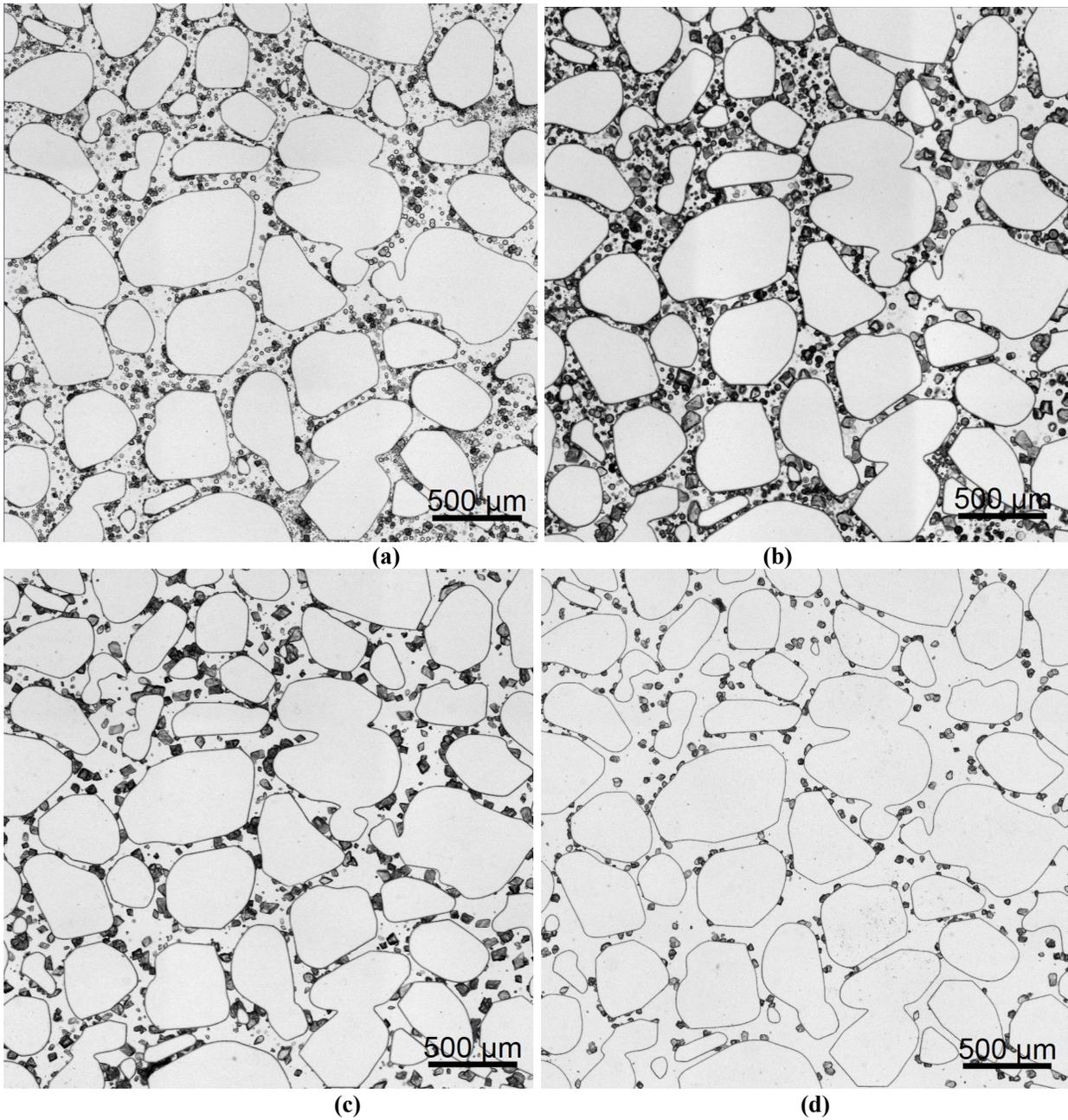

**Figure 11 3 mm by 3 mm microscope images taken at the central of the microfluidic chips taken at the completion of the MICP treatments. (a) 4°C, (b) 20°C, (c) 35°C, (d) 50°C**

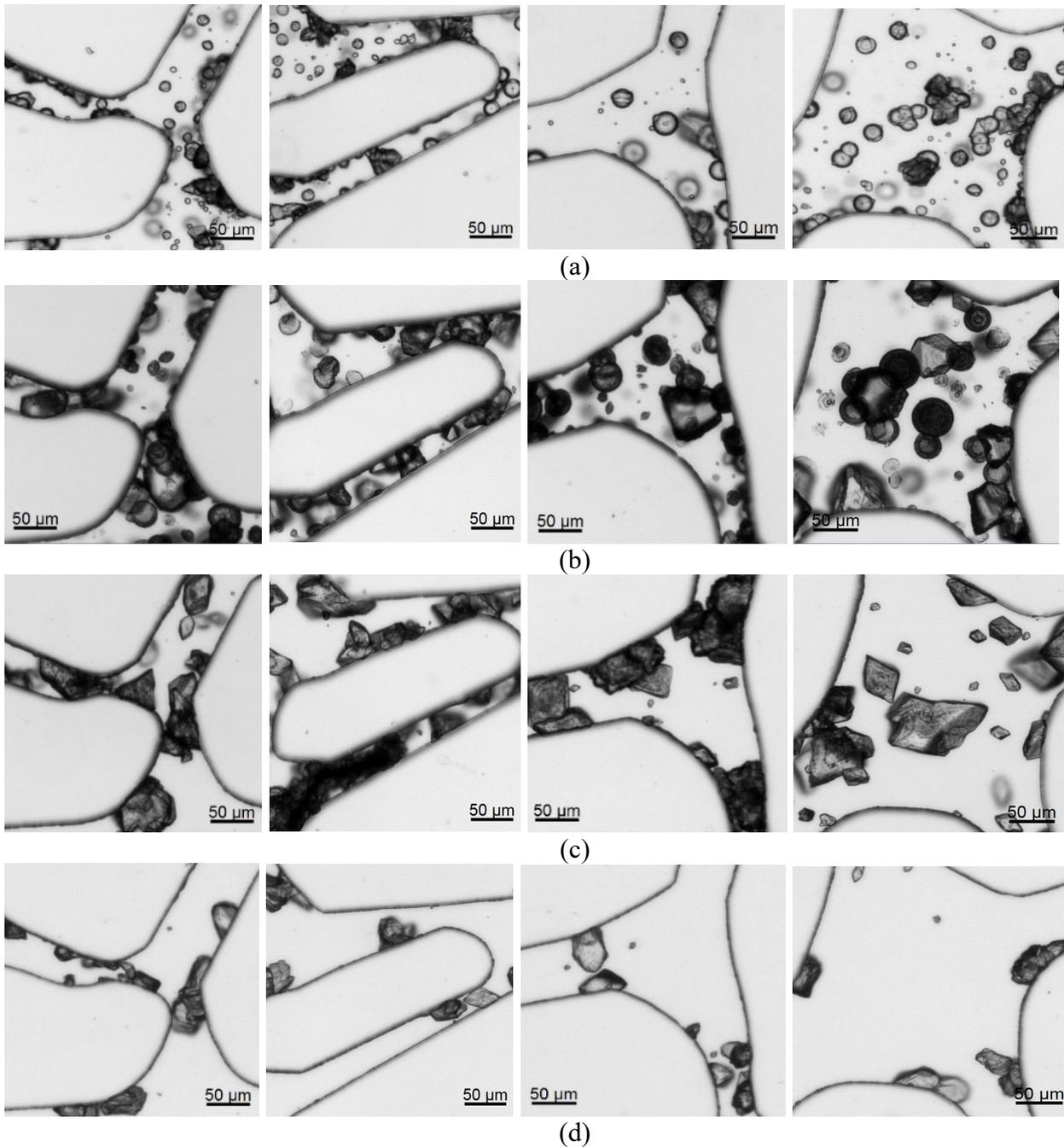

**Figure 12 magnified images at four pores in images shown in Figure 11. (a) 4°C, (b) 20°C, (c) 35°C, (d) 50°C**

**Optimization of MICP treatment procedures based on the findings**

This study has shown that, at high temperatures such as 50°C (Test 4), crystal growth rate should be high if the bacterial density and urease activity are constant (Figure 6c). However, due to the effects of high temperature on the reduction of both bacterial density and bacterial specific urease activity (Figure 4 a, c), the amount of $CaCO_3$ precipitated and chemical transformation efficiency of MICP were even lower than at 4°C (Figure 8 c, d). At 50°C, bacterial density and specific urease activity both decreased very quickly with time (Test 4, Figure 3a, c). Because crystals almost completed the precipitation by 6 hours after the first injection at 50°C (Test 4, Figure 6a), it is considered that reducing the injection interval from 48 hours to 6-18 hours (Test 5) and increasing bacterial injection number (Test 6) would increase the chemical transform efficiency of MICP at 50°C. The results in Figure 13 a, b show that reducing the injection interval to 6-18 hours from 48 hours helped to increase the total volume of crystals precipitated after the first two injections, but not afterwards. The total crystal volume after the fourth injection of cementation solution was higher in the case where additional injections of bacterial suspension were applied than in the other two cases, with $V_{crystals}/V_{pore}$ increasing linearly with each subsequent injection at a rate about 2.8%, per two injections (Figure 13b). The chemical transformation efficiency remained relatively high in the case where additional injections of bacterial suspension were applied, which is about 70% over the course of the ten injections, whereas for the other two cases, the chemical transformation efficiency decreased after each subsequent injection (Figure 13c). Figure 14 shows the images of the three cases at a larger scale (Figure 14a) and with more pores being shown (Figure 14b). The crystal precipitation patterns and characteristics are consistent with the images shown in Figure 13.

This study also found that the crystal growth rate is also low at a low temperature of 4°C. The reason for this is mainly because the precipitation rate of $CaCO_3$ is lower at lower temperatures (Figure 6a). Wang et al. (2021) found that, when temperature is constant, increasing bacterial density helped to increase the number of $CaCO_3$ crystal precipitated, but it did not affect the growth rate of the individual crystals. Therefore, higher bacterial density increases the total precipitation rate. Because the *in-situ* growth and attachment of bacteria at 4°C is relatively low (Figure 3), increasing the density of bacterial suspension injected would increase the precipitation rate of $CaCO_3$. However, Wang et al. (2021) found that, when bacterial density is higher, larger number of smaller crystals can be produced and the crystal dissolution-reprecipitation process takes longer. In this study, the bacterial density in the microfluidic chip at 4°C is lower than that at 20°C or 35°C, but the number of crystals is still much higher (Figure 8b). This is possible because lower temperature impairs the crystal phase transformation process (Figure 9 and 10). Based on the findings of Wang et al. (2021) and this study, the application of a higher bacterial density would increase the precipitation rate and chemical transform efficiency, but the crystals would be larger in amount and smaller in size. Therefore, compared to the high temperature condition which is easier to optimize by introducing larger numbers of bacterial injections, the optimization of MICP treatment procedures at low temperatures is more challenging and will therefore require further investigations.

Previous studies conducted at room temperature normally considered the initial bacterial activity (which affects the precipitation rate) when determining the injection interval for MICP treatment protocols (Al Qabany et al, 2012; Wang et al, 2019). However, the findings presented in the current study illustrate that, whilst bacterial activity may decrease quickly with time during the MICP treatment procedure, the initial bacterial activity cannot predict the precipitation rate of the whole process due to reasons such as temperature. In addition, the result at 50°C also indicates that the main reason why the crystals stopped growing after the 4[th] injection was because the bacterial activity was zero, with extra bacterial supply being required to maintain the required bacterial activity in the precipitation system.

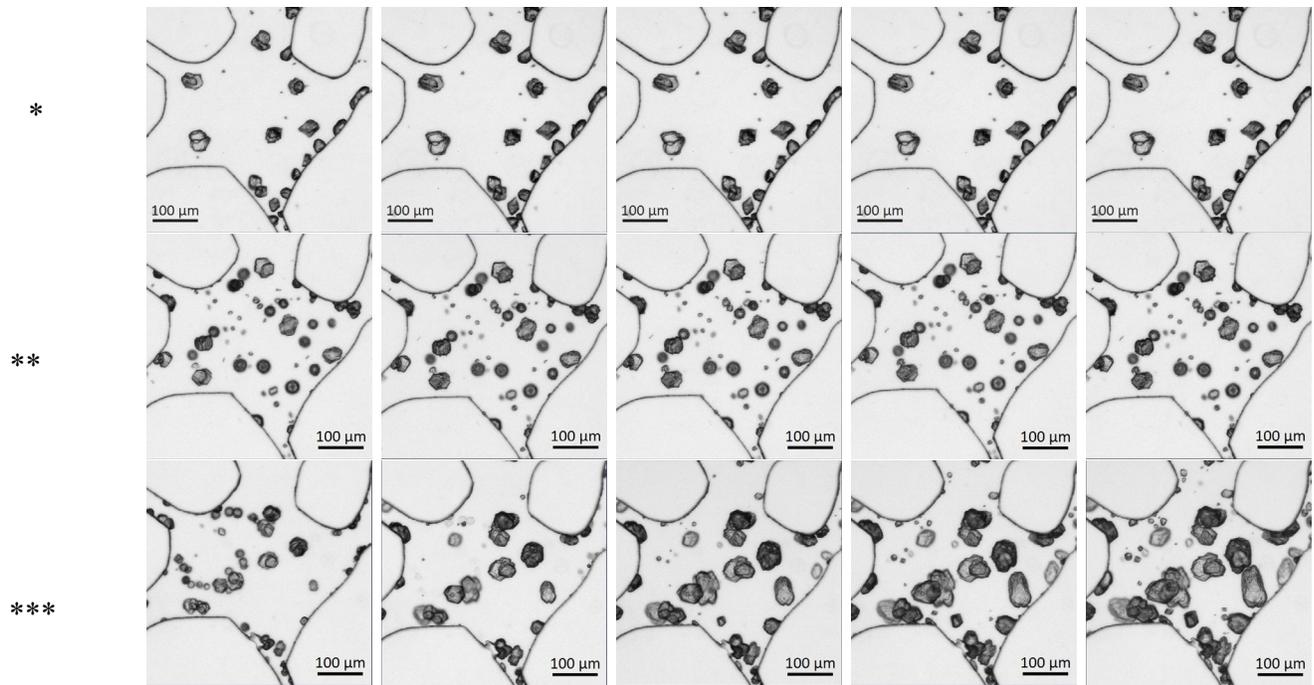

(a)

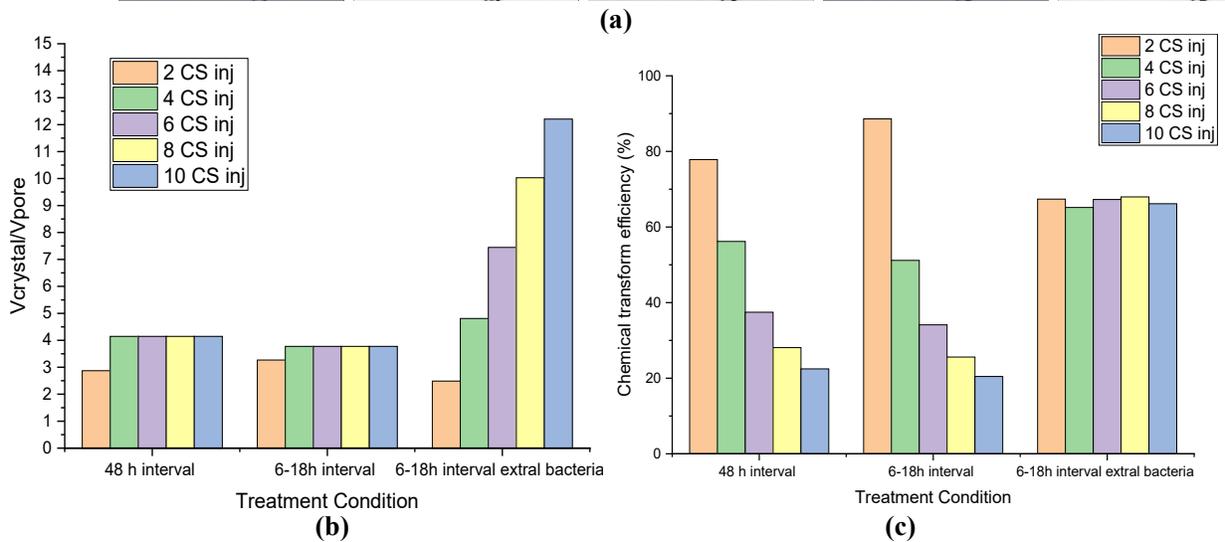

(b)                                           (c)

**Figure 13 Microscope images of one central pore of the microfluidic chips taken at the completion of the injection intervals of the second, fourth, eighth, and tenth injections of cementation solution for the sample treated by an injection interval of 48 hours (*), 6-18 hours (**) and 6-18 hours with bacterial injection before every two injections of cementation solution; (b) total crystal volume/Pore volume v.s. injection number of cementation solution; (c) chemical transform efficiency v.s. injection number of cementation solution**

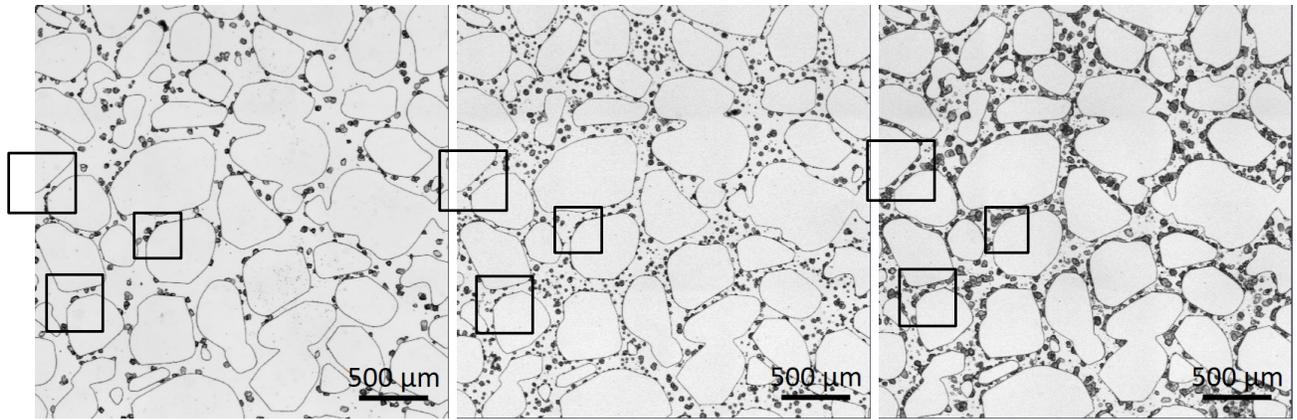

(a)

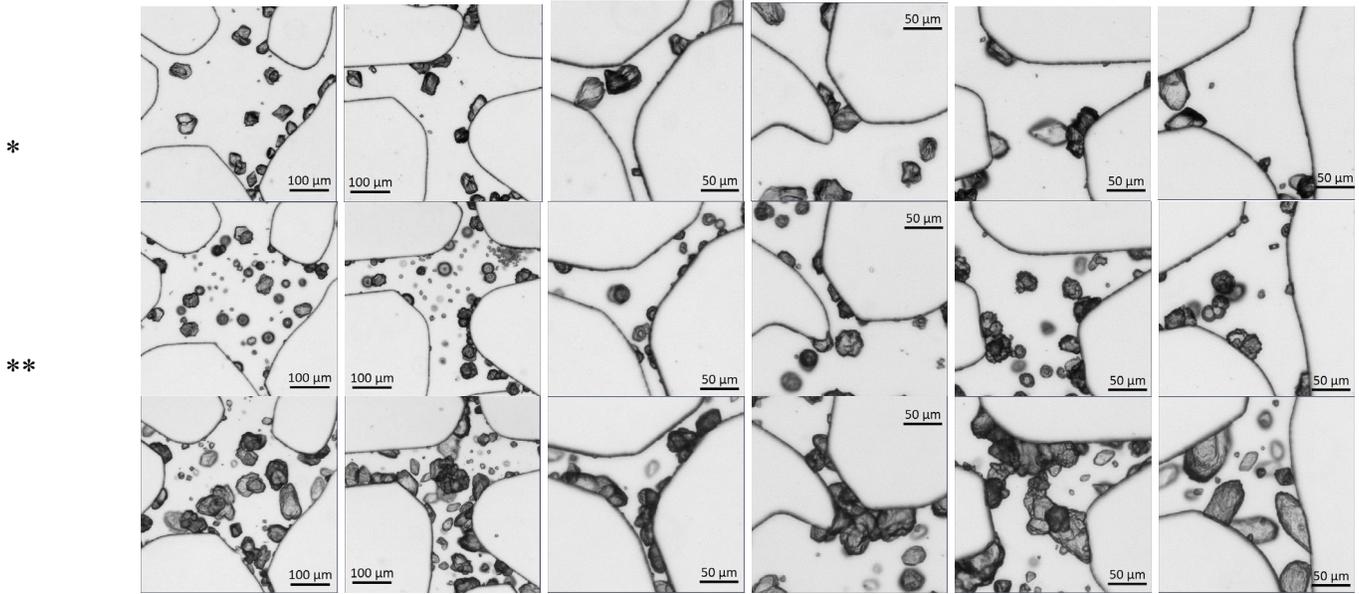

(b)

**Figure 14 (a) 3 mm by 3 mm microscope images taken at the central of the microfluidic chips taken at the completion of the MICP treatments at temperature of 50°C and an injection interval of 48 hours (left image), 6-18 hours (middle image) and 6-18 hours with extra injections of bacterial suspension before every other injections of cementation solution (right image); (b) magnified images at six pores in image (a).**

## CONCLUSIONS

In this study, bacterial activity batch tests and microfluidic chip experiments were conducted to investigate the effects of temperature on bacterial activity, bacterial number, crystal growth, crystal dissolution, crystal characteristics, and the chemical transform efficiency of MICP. The reasons for the low crystal growth rates at low and high temperatures are detailed and a method of optimizing MICP at high temperatures is proposed. These findings made from the microscope images have implications for better understanding of the MICP mechanisms at various temperatures and could be valuable for improving the MICP process for different applications. The main findings are as follows.

The effects of temperature on MICP crystal growth and efficiencies can be divided into the indirect and direct effects. The indirect ones refer to the effects of temperature on bacterial growth, attachment and ureolysis activity, which in turn influence the supersaturation condition of $CaCO_3$ precipitation. Direct one refers to the effect of temperature on $CaCO_3$ precipitation directly. The relatively low crystal precipitation rates and chemical efficiencies at 4°C are mainly affected by the low bacterial number and low precipitation rate of $CaCO_3$ at low temperature. The relatively high crystal precipitation rates and chemical efficiencies at 20°C are mainly due to the relatively higher number of bacterial cells after growth and detachment, and relatively higher overall crystals precipitation rate, which were not greatly affected by temperature. The higher initial crystal precipitation rate and chemical transform efficiencies at 35°C is the same as for 20°C. At 50°C, the reductions of crystal precipitation rate and chemical transform in a larger range of precipitation time (2-10 injections) are mainly because of the quick reduction rates in bacterial activity at 50°C.

The effects of temperature on $CaCO_3$ morphology are related to different crystal dissolution behaviour of $CaCO_3$ at different temperatures. The dissolution of vaterite and reprecipitation into larger vaterite or calcite follows the Ostwald law of crystal growth and a rise in temperature increases the rate of transformation. Therefore, high temperature conditions enhance the formation of calcite, whereas low temperature conditions enhance the formation of vaterite. This study also observed a 'memory-like' effect for the crystal-dissolution-reprecipitation process. When a new batch of cementation solution is injected, the crystals tend to reprecipitate where the crystals used to be during the previous injection before they dissolved. This is probably because the crystal lattice remained on the channel surface of the porous medium, which provide less energy for nucleation to occur compared to nuclear on clean surface.

Although initial precipitation rate increases exponentially with temperature increase, the precipitation efficiency of MICP at 50°C (Test 4) is very low because the high temperature reduces bacterial density and activity. At high temperature, therefore, increasing the injection number of bacterial suspension from 1 to 5 (Test 6) was found to be effective in increasing the $CaCO_3$ precipitation rate and chemical transform efficiency.

At low temperature, the low precipitation rate and chemical transformation efficiency are mainly because of the low $CaCO_3$ precipitation rate. The increasing bacterial density at low temperature may increase both precipitation rate and chemical transformation efficiency of MICP and the low temperature promotes the precipitation of a larger number of smaller crystals. However, the reduced $CaCO_3$ phase transformation rate would result in less effective crystals in bonding soil particles. The optimization of MICP for soil strength enhancement at low temperature is more challenging and requires further investigations.

AUTHOR INFORMATION
**Corresponding Author**
* E-mail: wangyz@sustech.edu.cn

**Author Contributions**
The manuscript was written through contributions of all authors. All authors have given approval to the final version of the manuscript.

**ACKNOWLEDGMENT**
Y.W. acknowledges the financial support of Natural Science Foundation of China (Grant No. 52171262, 42141003) and Science and Technology Innovation Committee of Shenzhen (Grant No. JCYJ20210324103812033) and Southern Marine Science and Engineering Guangdong Laboratory (Guangzhou, Grant No. K19313901) for conducting this study. The authors acknowledge Dr. Fedir Kiskin for proof reading the manuscript.

**ABBREVIATIONS**
MICP, Microbially-Induced Carbonate Precipitation; BS, bacterial suspension; CS, cementation solution; YE, Yeast Extract

Ferris F.G., Phoenix V., Fujita Y. (2004). Kinetics of calcite precipitation induced by ureolytic bacteria at 10 to 20 °C in artificial groundwater. Geochimica et Cosmochimica Acta, 68(8):1701–1710.

Gao, Z., Bian, L., Hu, Y., Wang, L., & Fan, J. (2007). Determination of soil temperature in an arid region. Journal of arid environments, 71(2), 157-168.

Green, F. H. W., Harding, R. J. (1980). Altitudinal gradients of soil temperatures in Europe. Transactions of the Institute of British Geographers, 5(2): 243-254.

Han, P., Geng, W., Li, M., Jia, S., Yin, J., and Xue R. (2021). Improvement of Biomineralization of Sporosarcina pasteurii as Biocementing Material for Concrete Repair by Atmospheric and Room Temperature Plasma Mutagenesis and Response Surface Methodology. J. Microbiol. Biotechnol. 31(9): 1311–1322

Jiang Ning-Jun, Liu Rui, Du Yan-Jun, Bi Yu-Zhang. (2019). Microbial induced carbonate precipitation for immobilizing Pb contaminants: Toxic effects on bacterial activity and immobilization efficiency. Science of the Total Environment, 672: 722–731. https://doi.org/10.1016/j.scitotenv.2019.03.294

Khan, Y. M., Munir, H., & Anwar, Z. (2019). Optimization of process variables for enhanced production of urease by indigenous Aspergillus niger strains through response surface methodology. Biocatalysis and Agricultural Biotechnology, 20(April), 101202.

Koh, C. A., Westacott, R. E., Zhang, W., Hirachand, K., Creek, J. L., & Soper, A. K. (2002). Mechanisms of gas hydrate formation and inhibition. *Fluid Phase Equilibria*, *194*, 143-151.

Kralj, D., Brečević, L., & Kontrec, J. (1997). Vaterite growth and dissolution in aqueous solution III. Kinetics of transformation. Journal of crystal growth, 177(3-4), 248-257.

Mitchell A C, Phillips A J, Hiebert R. Biofilm enhanced geologic sequestration of supercritical $CO_2$[J]. International Journal of Greenhouse Gas Control, 2009, **3**(1):90-99.

Mortensen B.M., Haber M.J., DeJong J.T. (2011) Effects of environmental factors on microbial induced calcium carbonate precipitation. Journal of Applied Microbiology, 111(2):338–349.

Nader Hataf & Alireza Baharifard. (2020). Reducing Soil Permeability Using Microbial Induced Carbonate Precipitation (MICP) Method: A Case Study of Shiraz Landfill Soil, Geomicrobiology Journal, 37:2, 147-158, DOI: 10.1080/01490451.2019.1678703

Nemati, M., Voordouw G. (2003). Modification of porous media permeability, using calcium carbonate produced enzymatically in situ. Enzyme and Microbial Technology 33: 635–642.

Peltzer, Edward T, Brewer, Peter G. (2000). Practical physical chemistry and empirical predictions of methane hydrate stability. In Natural Gas Hydrate (pp. 17-28). Springer, Dordrecht.

Peng Dinghua, Qiao Suyu, Luo Yao, Ma Hang, Zhang Lei, Hou Siyu, Wu Bin, Xu Heng. (2020). Performance of microbial induced carbonate precipitation for immobilizing Cd in water and soil. Journal of Hazardous Materials 400: 123116.

Peng, J., & Liu, Z. (2019). Influence of temperature on microbially induced calcium carbonate precipitation for soil treatment. PloS one, 14(6), e0218396.

Sagidullin, A. K., Stoporev, A. S., & Manakov, A. Y. (2019). Effect of temperature on the rate of methane hydrate nucleation in water-in-crude oil emulsion. Energy & Fuels, 33(4), 3155-3161.

Stocks-Fischer S., Galinat J.K., Bang S. (1999). Microbiological precipitation of $CaCO_3$. Soil Biological Biochemistry, 31(11):1563–1571.

Sun, X., Miao, L., Tong, T., & Wang, C. (2018). Study of the effect of temperature on microbially induced carbonate precipitation. Acta Geotechnica, 14(3), 627-638.

van Paassen, L. A., Ghose, Ranajit, van der Linden Thomas, J. M., van der Star Wouter, R. L., van Loosdrecht Mark, C. M. (2010). Quantifying Biomediated Ground Improvement by Ureolysis: Large-Scale Biogrout Experiment. Journal of Geotechnical and Geoenvironmental Engineering, 136(2): 1721-1728.27